\documentclass{article}

\usepackage{graphicx}
\usepackage{amsmath}

\def\prl#1#2#3{{\it Phys.\ Rev.\ Lett.} {\bf #1}, #2 (19#3)}
\newcommand{\prlb}[3]{{\it Phys.\ Rev.\ Lett.} {\bf #1}, #2 (20#3)}

\def\gsim{\;\raisebox{-.6ex}{$\stackrel{>}{\sim}$}\;}
\newcommand{\ra}{\rightarrow}
\newcommand{\beq}{\begin{equation}}
\newcommand{\eeq}{\end{equation}}
\newcommand{\Eq}[1]{Eq.~(\ref{eqM.#1})}

\begin{document}

\baselineskip 18pt

\begin{center}
{\Large {\bf Neutrino Intrinsic Properties: The Neutrino-Antineutrino Relation\footnote{To appear in Neutrino Physics, Proceedings of the Nobel Symposium 2004. Eds. L Bergstr\"{o}m, O. Botner, P. Carlson, P.O. Hulth, and T. Ohlsson} }} 
\vskip5mm  Boris Kayser\footnote{Email address: boris@fnal.gov}
\vskip3mm  Fermilab, MS 106, P.O. Box 500, Batavia IL 60510

\bigskip

\begin{abstract} 
Are neutrinos their own antiparticles? We explain why they very well might be. Then, after highlighting the fact that, to determine experimentally whether they are or not, one must overcome the smallness of neutrino masses, we discuss the one approach that nevertheless shows great promise. Finally, we turn to the consequences of neutrinos being their own antiparticles. These consequences include unusual electromagnetic properties, and manifestly CP-violating effects from ``Majorana'' phases that have no quark analogues.
\bigskip 
\end{abstract} 
\end{center} 

\section{Introduction}

The recent discovery that neutrinos have nonzero masses and mix implies that these particles must have very interesting intrinsic properties. At the Symposium, we discussed what is known, and what we would like to learn, about a number of these properties. In this written version of the talk, we would like to focus on one property: the relation between a neutrino and its antineutrino.

\section{ The Neutrino-Antineutrino Relation}

One of the most interesting questions about the intrinsic nature of
neutrinos, raised by the discovery of neutrino mass, is the question of
whether neutrinos are their own antiparticles. Is each neutrino mass
eigenstate $\nu_i$ identical to its antiparticle $\overline{\nu_i}$, or
distinct from it? If $\overline{\nu_i} = \nu_i$, we call the neutrinos
Majorana particles, while if $\overline{\nu_i} \neq \nu_i$, we call
them Dirac particles.

Of course, we know that the electron is distinct from its antiparticle,
the positron, because these two particles carry opposite electric charge. However, a neutrino carries no electric charge, and may not carry any other conserved charge-like quantum number. It might be thought that there is a conserved lepton number $L$, defined by 
\beq
L(\nu) = L(\ell^-) = -L(\bar{\nu}) = -L(\ell^+) = 1 ~~,
\label{eqM.1}
\eeq
that distinguishes neutrinos $\nu$ and charged leptons $\ell^-$ on the one hand from antineutrinos $\bar{\nu}$ and anti-charged leptons $\ell^+$ on the other hand. However, there is no evidence that such a conserved quantum number exists. If it does not exist, then nothing distinguishes $\overline{\nu_i}$ from $\nu_i$. The neutrinos are then Majorana particles, identical to their antiparticles.

Many theorists believe that, indeed, the lepton number $L$ defined by \Eq{1} is not conserved. One reason for this belief is the nature of the very successful Standard Model (SM). As originally proposed, the SM conserves $L$. However, it contains no neutrino masses. Nor does it contain any chirally right-handed neutrino fields, $\nu_R$, but only left-handed ones, $\nu_L$.

Now that we know neutrinos have masses, we must extend the SM to accommodate them. Suppose that we try to do this in a manner that will preserve the conservation of $L$. Then, for a neutrino $\nu$, we add to the SM Lagrangian a ``Dirac mass term'' of the form
\beq
\mathcal{L} = -m_D \overline{\nu_L} \nu_R + h.c.~~ .
\label{eqM.2}
\eeq
Here, $m_D$ is a constant, and $\nu_R$ is a right-handed neutrino field that we were obliged to add to the SM in order to construct the Dirac mass term. A Dirac mass term does not mix neutrinos and antineutrinos, so it conserves $L$.

In the SM, left-handed fermion fields belong to electroweak-isospin doublets, but right-handed ones are isospin singlets. In particular, $\nu_R$ will carry no electroweak isospin. Thus, once $\nu_R$ is present, all the SM principles, including electroweak-isospin conservation, allow the occurrence of the ``Majorana mass term''
\beq
\mathcal{L}_M = -m_M \overline{\nu_R^c} \nu_R + h.c.~~ .
\label{eqM.3}
\eeq
Here, $m_M$ is a constant, and $\nu_R^c$ is the charge conjugate of $\nu_R$. Like $\nu_R$, $\nu_R^c$ carries no electroweak isospin. Thus, $\mathcal{L}_M$ is electroweak-isospin conserving, as required. However, any Majorana mass term of this form converts a $\nu$ into a $\bar{\nu}$, or a $\bar{\nu}$ into a $\nu$. Thus, it does not conserve $L$.

If we insist that the SM, extended to accommodate neutrino masses, remain $L$ conserving, then, of course, Majorana mass terms are forbidden. However, if we do not impose $L$ conservation by hand, but require only the general SM principles, such as electroweak-isospin conservation and renormalizability, then Majorana mass terms such as the one in \Eq{3} are allowed. It is then very natural to expect that they are present in nature, so that $L$ is not conserved and the neutrinos are Majorana particles.

The most popular explanation of why neutrinos are so light is the see-saw mechanism, discussed at this Symposium by P. Ramond. This mechanism includes Majorana mass terms. Hence, it predicts that $L$ is not conserved and that neutrinos are Majorana particles.

\begin{figure}[!hbp]
\begin{center}
\includegraphics[clip=true,scale=0.5]{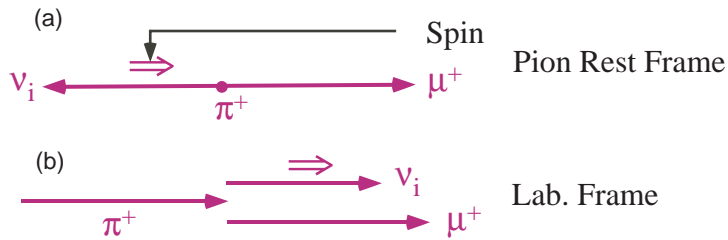} \bigskip
\caption{A pion decay into a muon and a neutrino mass eigenstate. (a) The decay as seen in the pion rest frame. (b) The same decay as seen in the laboratory frame when the pion is moving faster in that frame than the neutrino is moving in the pion rest frame.} 
\label{f1} 
\end{center}
\end{figure}

How can we test whether neutrinos are their own antiparticles? Let us first discuss an approach \cite{r1} that would not work, but that clearly illustrates why most approaches would not work. Suppose that a neutrino mass eigenstate $\nu_i$ is produced in $\pi^+$ decay, as depicted in Fig.\ref{f1}(a). We assume, as shown there, that the $\nu_i$ is emitted to the left in the $\pi^+$ rest frame. 
Owing to the left-handed character of the SM weak interactions, the $\nu_i$ will have left-handed (LH) helicity, so that its spin vector will point to the right, as shown. Now imagine that the parent $\pi^+$ is moving to the right in the laboratory frame at a speed greater that that of the $\nu_i$ in the $\pi^+$ rest frame. Then, as indicated in Fig.~\ref{f1}(b), the $\nu_i$ will be moving to the right in the laboratory frame. Moreover, its spin vector will be pointing to the right in this frame, just as it was in the $\pi^+$ rest frame. Thus, in the laboratory frame, the $\nu_i$ will have {\em right-handed} (RH) helicity.

Now, the statement that a neutrino $\nu_i$ is its own antiparticle means that, {\em for a given helicity $h$},
\beq
\overline{\nu_i}(h) = \nu_i(h) ~~.
\label{eqM.4}
\eeq
It is very difficult to test whether this equality holds in nature because, thanks to the left-handed character of the weak interactions, weak processes produce neutrinos that are almost exclusively of LH helicity, but antineutrinos that are almost exclusively of RH helicity. Thus, the comparison called for by \Eq{4} cannot be made. 
However, in our hypothetical experiment with $\pi^+$ decays, we have created a sample of $\nu_i$ that are tagged as neutrinos, rather than antineutrinos, because they are produced together with $\mu^+$ particles, but which have RH helicity in the laboratory frame because of the very rapid motion of their parent pions. Are these RH $\nu_i$ identical to RH $\overline{\nu_i}$, in conformity with \Eq{4}? That is, do these RH $\nu_i$ interact as RH $\overline{\nu_i}$ do? A  RH $\overline{\nu_i}$  enjoys the SM charged-current interaction. When this particle strikes a target at rest, this interaction allows it to produce a charged antilepton, $\ell^+$. Can the RH $\nu_i$ in our hypothetical $\pi^+$ decay experiment do the same?

Unfortunately, we cannot turn our hypothetical experiment into a real one to answer this question. The reason is that neutrino masses are much too small. Suppose, for example, that $\nu_i$ has a mass $m_i$ around 0.05 eV, a value suggested by atmospheric neutrino oscillation data. A $\pi$ of mass $m_\pi$, moving faster in the laboratory frame than its daughter $\nu_i$ moves in the $\pi$ rest frame, must have a laboratory energy $E_\pi$(Lab) obeying
\beq
\frac{E_\pi({\rm Lab})}{m_\pi} > \frac{E_{\nu_i}(\pi~{\rm Rest \; Frame})}{m_i} ~~ .
\label{eqM.6}
\eeq
For $m_i \sim$ 0.05 eV, this requires that $E_\pi({\rm Lab}) \gsim 10^5$ TeV. Furthermore, a $\nu_i$ from $\pi$ decay will not get its helicity reversed by the forward motion of the $\pi$ unless, in the $\pi$ rest frame, its angle of emission is within $\sim m_i / E_{\nu_i} (\pi$ Rest Frame) of dead backward with respect to the $\pi$ beam direction \cite{r1}. Thus, the fraction of all $\pi$-decay $\nu_i$ that get their helicity reversed is only $\sim [m_i / E_{\nu_i} (\pi$ Rest Frame)]$^2$. For $m_i \sim$ 0.05 eV, this fraction is 10$^{-18}$. Obviously, our hypothetical experiment is completely impossible in practice.

This practical impossibility illustrates a very general point: We are trying to find a way to demonstrate that $\overline{\nu_i} = \nu_i$, or, equivalently, that lepton number $L$ is not conserved. In this effort, we are assuming that the interactions of neutrinos are correctly described by the SM. Now, the SM interactions conserve $L$, so the $L$ nonconservation that we seek can only come from Majorana neutrino mass terms. Thus, it must vanish when the neutrino masses vanish. Hence, any attempt to demonstrate that $\overline{\nu_i} = \nu_i$ or that $L$ is not conserved will be challenged by the smallness of neutrino masses. Our experiment based on $\pi$ decay cannot meet this challenge. Indeed, the only approach that shows considerable promise of being able to meet it is the search for neutrinoless double beta decay.

Neutrinoless double beta decay $(0\nu\beta\beta)$ is the reaction Nucl $\rightarrow$ Nucl$^\prime + e^- + e^-$, in which one nucleus decays into another plus two electrons and nothing else. Manifestly, this reaction would not conserve $L$. Thus, observing it at any nonzero level \cite{r2} would establish that neutrinos are identical to their antiparticles. Like any $L$-nonconserving process, $0\nu\beta\beta$ is suppressed by the smallness of neutrino masses. However, if we choose as our parent nucleus one that cannot decay by $\alpha$ or single $\beta$ emission, and wait long enough, we might see it decay by $0\nu\beta\beta$ emission. To be sure, any nucleus that can decay in this $L$-nonconserving way can also decay via the $L$-conserving process Nucl $\rightarrow$ Nucl$^\prime + e^- + e^- + \bar{\nu} + \bar{\nu}$. However, this two-neutrino double beta decay is phase-space suppressed, giving the neutrinoless mode a chance to be observed.

\begin{figure}[!hbp]
\begin{center}
\includegraphics[clip=true,scale=0.5]{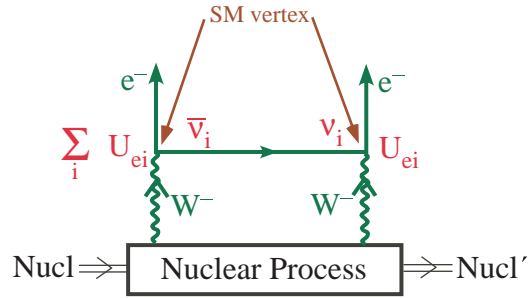} \bigskip
\caption{The neutrino-exchange mechanism for $0\nu\beta\beta$.} 
\label{f2} 
\end{center}
\end{figure}

The dominant mechanism for $0\nu\beta\beta$ is expected to be the neutrino-exchange diagram in Fig.~\ref{f2}, in which one or another of the neutrino mass eigenstates $\nu_i$ is exchanged.
The neutrino-electron-W-boson vertices in this diagram are assumed to be SM weak vertices, which conserve $L$. Thus, if $\overline{\nu_i}$ is distinct from $\nu_i$, the exchanged particle emitted by the leptonic weak vertex on the left side of the diagram must be a $\overline{\nu_i}$. But when this same exchanged particle is absorbed by the leptonic weak vertex on the right side of the diagram, it must be a $\nu_i$. Thus, this diagram does not exist if $\overline{\nu_i} \neq \nu_i$, but only if $\overline{\nu_i} = \nu_i$.

Apart from an overall coupling strength, the amplitude for a $\nu_i$ to create a charged lepton of ``flavor'' $\alpha$ at a SM weak vertex is $U_{\alpha i}$, where $U$ is the unitary leptonic mixing matrix.  Hence, there is a factor of $U_{ei}$ at each of the leptonic weak vertices in Fig.~\ref{f2}. As indicated in that figure, the amplitude for $0\nu\beta\beta$, Amp [$0\nu\beta\beta$], is a coherent sum over the contributions of the different $\nu_i$.Just as if it had been born in an $e^-$-producing $\beta$ decay, the exchanged $\nu_i$ in Fig.~\ref{f2} is emitted in a state which is almost totally of right-handed helicity, but which contains a small piece, of order $m_i / E_{\nu_i}$, having left-handed helicity. Here, as before, $m_i$ is the mass of $\nu_i$, and $E_{\nu_i}$ is its energy. When the exchanged $\nu_i$ is absorbed, the absorbing SM left-handed current can only absorb its left-handed component without further suppression. Since this component is $\mathcal{O}[m_i / E_{\nu_i}]$, the contribution of $\nu_i$ exchange to $0\nu\beta\beta$ is proportional to $m_i$. Hence, recalling the two factors of $U_{ei}$ in Fig.~\ref{f2}, and summing over all the $\nu_i$ contributions,
\beq
{\rm Amp}[0\nu\beta\beta] \propto \left| \sum_i m_i U_{ei}^2 \right| \equiv m_{\beta\beta} ~~.
\label{eqM.7}
\eeq
The quantity $m_{\beta\beta}$ is known as the effective Majorana neutrino mass for neutrinoless double beta decay.

As we have stressed, if neutrino interactions are governed by the SM, then any $L$ nonconservation in nature must vanish with the neutrino masses. \Eq{7} makes this vanishing explicit for the case of $0\nu\beta\beta$.

The fact that Amp [$0\nu\beta\beta$] depends on neutrino masses means that a measurement of the rate for $0\nu\beta\beta$ would provide information on these masses. This was discussed at this Symposium by S. Petcov.

\section{Consequences of Majorana Character}

Suppose that, as many theorists suspect, neutrinos are indeed their own antiparticles. What would be the physical consequences?

\subsection{Electromagnetic Properties}\label{C.1}
If neutrinos are their own antiparticles, then they are not only electrically neutral, but also devoid of any internal charge distribution. To see why, suppose, for example, that a neutrino has a charge distribution consisting of a positively-charged core surrounded by a compensating negatively-charged shell. The CPT-mirror image of this neutrino would have a negatively-charged core surrounded by a positively-charged shell. But if the neutrino is its own antiparticle, then  (apart from a reversal of its spin), it must be identical to its CPT-mirror image. Thus a neutrino that is its own antiparticle cannot have a charge distribution.

A Majorana neutrino cannot have a magnetic or electric dipole moment either. From CPT invariance it follows that a fermion and its antiparticle have equal and opposite dipole moments. But a Majorana neutrino is its own antiparticle. Hence, it can have no dipole moments.

Both Majorana and Dirac neutrinos can have {\em transition} dipole moments connecting one neutrino mass eigenstate to another. Such moments make possible radiative decays such as $\nu_2 \ra \nu_1 + \gamma$. However, in the SM, extended to included neutrino mass, such decays are exceedingly slow \cite{r3}. For example, if the mass of $\nu_2$ is $\sim$0.05 eV and $\nu_1$ is massless, the lifetime for $\nu_2 \ra \nu_1 + \gamma$ is $\sim$10$^{49}$ yr. Visibly fast neutrino decays, electromagnetic or non-electromagnetic, would entail physics beyond even the extended SM. It is interesting that for some years, the Super-Kamiokande atmospheric $\nu_\mu$ disappearance data could be explained either in terms of neutrino oscillation or in terms of neutrino decay \cite{r4}. However, the decay hypothesis is now excluded by Super-Kamiokande results, including their $L/E$ (Distance/Energy) analysis \cite{r5}.

\subsection{CP Violation in Neutrino Oscillation}\label{C.2}
\begin{figure}[!htp]
\begin{center} 
\includegraphics[clip=true,scale=0.5]{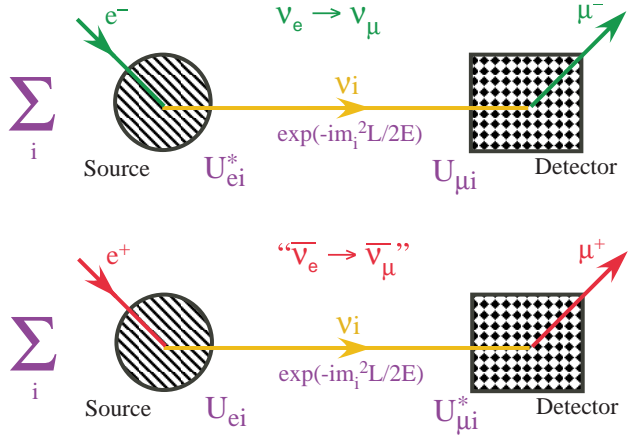}
\caption{The amplitudes for the oscillations $\nu_e \ra \nu_\mu$ and ``$\overline{\nu_e} \ra \overline{\nu_\mu}$'', showing the charged leptons that define these processes. For each process, we show the elements of the mixing matrix $U$ that appear in the amplitude where the neutrino is produced. and where it is detected. We also show the amplitude $\exp{(-m_i^2 L/2E)}$ for each mass eigenstate $\nu_i$ to propagate for a distance $L$ with energy $E$. The figure assumes that $\overline{\nu_i} = \nu_i$, but nothing would change if in fact $\overline{\nu_i} \neq \nu_i$, so that the propagating neutrino in  ``$\overline{\nu_e} \ra \overline{\nu_\mu}$'' is a $\overline{\nu_i}$ rather than a $\nu_i$.} 
\label{f3} 
\end{center}
\end{figure}

Neutrino oscillation is completely insensitive to whether neutrinos are of Dirac or Majorana character. In particular, CP violation in neutrino oscillation is completely insensitive to this question. At first sight, this might seem strange. After all, CP violation in neutrino oscillation would manifest itself as a difference between the probabilities for the CP-mirror-image oscillations $\nu_\alpha \ra \nu_\beta$ and ``$\overline{\nu_\alpha} \ra \overline{\nu_\beta}$'', where $a,\:\beta = e,\:\mu,\: {\rm or}\;\tau$. If neutrinos are Majorana particles, so that each mass eigenstate $\nu_i$ is identical to its antiparticle $\overline{\nu_i}$, is ``$\overline{\nu_\alpha} \ra \overline{\nu_\beta}$'' a different process from $\nu_\alpha \ra \nu_\beta$?
Indeed it is, as illustrated in Fig.~\ref{f3} for the example of $\nu_e \ra \nu_\mu$ and ``$\overline{\nu_e} \ra \overline{\nu_\mu}$''. 
A 
s shown in Fig.~\ref{f3}, each of these two processes is defined in practice by the flavors and {\em the signs of the charges} of the charged leptons at the source and detection ends of the neutrino beamline. When one goes from $\nu_e \ra \nu_\mu$ to ``$\overline{\nu_e} \ra \overline{\nu_\mu}$'', the signs of these charges reverse, so that $\nu_e \ra \nu_\mu$ and ``$\overline{\nu_e} \ra \overline{\nu_\mu}$'' are different processes, and they can have different probabilities, even when $\overline{\nu_i} = \nu_i$. The amplitude for each process is a coherent sum over the contributions of the various $\nu_i$, as shown in Fig.~\ref{f3}. Irrespective of whether $\overline{\nu_i} = \nu_i$, as assumed in Fig.~\ref{f3}, or not, the contribution of each $\nu_i$ to each process involves the factors indicated in this figure. Clearly, if the mixing matrix $U$ contains a CP-violating phase, the probabilities for $\nu_e \ra \nu_\mu$ and ``$\overline{\nu_e} \ra \overline{\nu_\mu}$'' can differ, even when $\overline{\nu_i} = \nu_i$.

\subsection{Majorana CP-Violating Phases}\label{C.3}

The $3 \times 3$ quark mixing matrix contains just one CP-violating phase. However, if neutrinos are Majorana particles, then the $3 \times 3$ leptonic mixing matrix may contain three CP-violating phases \cite{r6}. The two extra phases, $\xi_1$ and $\xi_2$, are called Majorana phases. The phase $\xi_i$ is associated with the neutrino mass eigenstate $\nu_i$. If, in the absence of the Majorana phases, the leptonic mixing matrix is $U^0$, then in their presence it is $U$, where
\beq
U_{\alpha i} = U_{\alpha i}^0 e^{i\frac{\xi_i}{2}} \;\;  ; {\rm all} \; \alpha ~~.
\label{eqM.8}
\eeq

Majorana phases have physical consequences only in processes that involve a violation of lepton number $L$ \cite{r7}. They do not affect neutrino oscillation, but do influence $0\nu\beta\beta$. Indeed, their influence on $0\nu\beta\beta$ is obvious from \Eq{7} for $m_{\beta\beta}$, the effective Majorana neutrino mass for $0\nu\beta\beta$. Clearly, $m_{\beta\beta}$ depends on the relative phase of $U^2_{e1}$ and $U^2_{e2}$; that is, on $\xi_1 - \xi_2$.

The ``Dirac'' CP-violating phase in the quark mixing matrix can lead to manifest CP violation, by which we mean a manifestly CP-violating inequality between the rate for some process, and the rate for its CP-mirror image. Can Majorana phases cause such a manifestly CP-violating inequality too? In theory, they can cause it in heavy neutrino decay in the early universe, leading to the excess of matter over antimatter that we see in the universe today (see Sec.~\ref{D}). This phenomenon, known as leptogenesis, was discussed at this Symposium by T. Yanagida. However, we wish to ask whether Majorana phases can cause CP-violating rate inequalities in present-day processes. The answer is yes \cite{r8}, although these inequalities occur only in processes that would be extremely difficult to observe. An example is $e^+ W^- \ra \nu \ra \mu^- W^+$, and its CP-mirror image, $e^- W^+ \ra \nu \ra \mu^+ W^-$. Here, the intermediate Majorana neutrino propagates down a beamline, as in a neutrino oscillation experiment, and the $W$ bosons are attached to nucleons at the neutrino source and at the detector.

The amplitude for $e^+ W^- \ra \nu \ra \mu^- W^+$ is given by
\beq
{\rm Amp}\, [e^+ W^- \ra \nu \ra \mu^- W^+] = S \sum_i U_{ei} U_{\mu i} \frac{m_i}{E} e^{-im_i^2 \frac{L}{2E} } ~~ .
\label{eqM.9}
\eeq
Here, the $i$'th term is the contribution of an intermediate $\nu_i$. $S$ is a kinematical factor, $m_i$ is the mass of $\nu_i$, $L$ is the distance traveled by the neutrino, and $E$ is its energy. The factor $\exp [-im_i^2\, L/2E]$ is the same neutrino propagator that occurs in ordinary neutrino oscillation. The factor $m_i / E$ is a suppression factor reflecting the fact that the incoming $e^+$ leads to a neutrino in a dominantly right-handed helicity state, whereas the left-handed current that produces the outgoing $\mu^-$ would prefer that the neutrino be in a left-handed state. It is this suppression factor that makes this process so hard to observe.

The amplitude for $e^- W^+ \ra \nu \ra \mu^+ W^-$ is given by
\beq
{\rm Amp}\, [e^- W^+ \ra \nu \ra \mu^+ W^-] = S \sum_i U^*_{ei} U^*_{\mu i} \frac{m_i}{E} e^{-im_i^2\frac{L}{2E} } ~~ .
\label{eqM.10}
\eeq
Comparing this with Amp [$e^+ W^- \ra \nu \ra \mu^- W^+$], and consulting \Eq{8}, we see that Majorana phases in the $U$ matrix will indeed lead to a CP-violating rate inequality. In fact, this would even be true if there were only two generations of leptons. To see this, let us suppose that there are only two generations: two charged leptons $e$ and $\mu$, and two neutrinos, $\nu_1$ and $\nu_2$. Then the most general unitary mixing matrix is
\beq
\begin{array}{ccc} 
     &    \hspace{.5cm} \nu_1 \hspace{.5cm}  \nu_2  &     \smallskip  \\
     U =  & \left[   \begin{array}{cc}
         \phantom{-}c\, e^{i\frac{\xi}{2}} & s  \\
                                 -s\, e^{i\frac{\xi}{2}} & c  \end{array} \right]
  & \begin{array}{c}   e  \\  \mu  \end{array} ~~ ,
\end{array}
\label{eqM.11}
\eeq
where $c$ and $s$ are, respectively, the cosine and sine of a mixing angle $\theta, \;\xi$ is a Majorana phase, and the symbols outside of the matrix label the columns and rows. Using this mixing matrix in \Eq{9} and squaring, we find that the rate $\Gamma$ for $e^+ W^- \ra \nu \ra \mu^- W^+$ is given by \cite{r9}
\beq
\Gamma [e^+ W^- \ra \nu \ra \mu^- W^+] = K \frac{\sin^2 2\theta}{4E^2} [m^2_1 + m^2_2 - 2m_1 m_2 \cos(\Delta m^2 \frac{L}{2E} - \xi)] ~~.
\label{eqM.12}
\eeq
Here, $K = |S|^2$, and $\Delta m^2 = m^2_2 - m^2_1$. Using the same mixing matrix in \Eq{10}, we find that 
\beq
\Gamma [e^- W^+ \ra \nu \ra \mu^+ W^-] = K \frac{\sin^2 2\theta}{4E^2} [m^2_1 + m^2_2 - 2m_1 m_2 \cos(\Delta m^2 \frac{L}{2E} + \xi)] ~~.
\label{eqM.13}
\eeq
We see that if the Majorana phase $\xi$ is present, the rates for the CP-mirror-image processes $e^+ W^- \ra \nu \ra \mu^- W^+$ and $e^- W^+ \ra \nu \ra \mu^+ W^-$ differ.

When compared with behavior familiar from the quark sector, the rates of Eqs.~(\ref{eqM.12}) and (\ref{eqM.13}) have a surprising feature. One may think of the quark mixing matrix as indicating the combination of d, s, and b quarks that couples to a particular up-type quark (such as the top quark), or, equivalently, the combination of u, c, and t quarks that couples to a particular down-type quark (such as the strange quark). Thus, if the masses of u, c, and t were equal, the quark mixing matrix would lose its meaning and all of its physical consequences, because we could no longer distinguish u, c, and t from one another. Similarly, the mixing matrix would lose its meaning and consequences if the masses of d, s, and b were equal. But what happens to the rates of  Eqs.~(\ref{eqM.12}) and (\ref{eqM.13}) when the masses of $\nu_1$ and $\nu_2$, $m_1$ and $m_2$, are equal? With $m_1 = m_2 \equiv m$, these rates become
\beq
\Gamma [e^+ W^- \ra \nu \ra \mu^- W^+] = \Gamma [e^- W^+ \ra \nu \ra \mu^+ W^-] = K \sin^2 2\theta \frac{m^2}{E^2} \sin^2 \frac{\xi}{2} ~~ .
\label{eqM.14}
\eeq
Evidently, if the Majorana phase $\xi$ is present, so that these rates do not vanish, the mixing angle $\theta$ in the leptonic mixing matrix still has physical consequences. To understand why this can happen, we note that through a change of phase conventions, the phase $\xi$ that is associated with neutrino $\nu_1$ in the mixing matrix $U$ of \Eq{11} can be removed from the $U$ matrix and attached to the mass of $\nu_1$ \cite{r8}. The mass of $\nu_1$ then becomes a complex quantity, $m_1 e^{i\xi}$. But the mass of $\nu_2$, a neutrino with no associated Majorana phase, is real. Thus, there is still a distinction between $\nu_1$ and $\nu_2$, even when their masses are of equal size. Hence, the leptonic mixing matrix can still have meaning and physical consequences.

\section{A concluding question}\label{D}

The exploration of the possible effects of Majorana phases raises a question: Why are there three generations of quarks and leptons?

It is well known that the presently observed preponderance of matter over antimatter in the universe could not have developed without a violation of CP sometime in the early universe. Once, it had been thought that this violation of CP came from the CP-violating phase in the quark mixing matrix. Had this been the case, we could have argued that---
\begin{itemize}
\item[a)] The phase in the quark mixing matrix cannot produce CP violation unless there are at least three generations.
\item[b)] Without this CP violation, there would not be a preponderance of matter over antimatter in the universe.
\item[c)] Without this preponderance, we would not be here to ask why there are three generations.
\end{itemize}

However, it is now well established that the phase in the quark mixing matrix could not have produced nearly enough CP violation to explain the present excess of matter over antimatter. As a result, there is a lot of interest in the appealing possibility that this excess resulted from leptogenesis. In the see-saw mechanism, each light neutrino $\nu$ is accompanied by a very heavy neutrino $N$. Both the light neutrino and its heavy see-saw partner are Majorana particles. In leptogenesis \cite{r10}, discussed at this Symposium by T. Yanagida, there is a CP violation, coming from Majorana phases, in the decays of the heavy neutrinos $N$ in the early universe. This CP violation leads to unequal rates for the leptonic decays $N \ra \ell^+ + {\rm Higgs}^-$ and $N \ra \ell^- + {\rm Higgs}^+$. These unequal rates lead, in turn, to a universe with unequal amounts of matter and antimatter, as observed.

In the two-generation example described in Sec.~\ref{C.3}, we saw that Majorana phases can already produce manifestly CP-violating rate inequalities, such as the one required for leptogenesis, when there are only {\em two} generations. So why are there {\em three}?

\bigskip\noindent {\large \bf Acknowledgements}

This work was supported by the U.S. Department of Energy.

\end{document}